\documentclass[conference]{IEEEtran}
\IEEEoverridecommandlockouts
\usepackage{cite}
\usepackage{amsmath,amssymb,amsfonts}
\usepackage{algorithmic}
\usepackage{graphicx}
\usepackage{textcomp}
\usepackage{xcolor}
\def\BibTeX{{\rm B\kern-.05em{\sc i\kern-.025em b}\kern-.08em
    T\kern-.1667em\lower.7ex\hbox{E}\kern-.125emX}}
\begin{document}

\title{An Abstract View on the De-anonymization Process
\thanks{\hrule height 0.5pt \vspace{4pt} This research was supported by the Safe-DEED project (HORIZON 2020, Grant Agreement No.: 825225), funded by the European Commission.}
}

\author{\IEEEauthorblockN{Alexandros Bampoulidis, Mihai Lupu}\IEEEauthorblockA{Research Studios Austria - Studio Data Science \\ Vienna, Austria \\ \{name.surname@researchstudio.at\}}}

\maketitle

\begin{abstract}
Over the recent years, the availability of datasets containing personal, but anonymized information has been continuously increasing.
Extensive research has revealed that such datasets are vulnerable to privacy breaches:
being able to reveal sensitive information about individuals through de-anonymization methods.
Here, we provide a taxonomy of the research in de-anonymization.
%In this short survey paper, we provide our abstract view on the de-anonymization process, targeted to data providers.
\end{abstract}


\section{Introduction}
De-anonymization is not as hard a task as one might think.
A common misconception is that removing all explicit identifiers such as name, address and phone number from a dataset, makes the data anonymous.
Several approaches described in the de-anonymization literature have proven this belief wrong.
The continuously increasing availability of datasets, either private or public, that can be used as auxiliary data in de-anonymization techniques and the fact that there is a privacy breach when even a single individual is identified from a dataset, make anonymization of data a difficult challenge to address, while making de-anonymization easier.

This short paper provides an abstract view of data de-anonymization and raises the awareness of data providers on this topic.
Our survey differs from the three other existing surveys on de-anonymization \cite{AlAzizy2015ALS,Ding2010,10.1371/journal.pone.0028071,Ji2017} in providing a more abstract view on this topic and being targeted at data providers.

In the next section, we present our view on the de-anonymization along with references to scientific publications.
Note that due to the space limitation, we only cite the most important and representative publications of their kind.
Finally, we provide a short discussion and implications for data providers on the de-anonymization process.

\section{De-anonymization Process}
We identify three elements in the de-anonymization process:
purpose, data (target and auxiliary), and approach.
\subsection{Purpose}
Purpose refers to why one would perform a de-anonymization attack on a dataset.
We identify the three following purposes:

\subsubsection{Proof of Concept}
An attack of such purpose aims in revealing whether a dataset is de-anonymizable or that it cannot be de-anonymized with reasonable effort.
This is an important action for data providers to take before releasing their dataset(s), keeping in mind that the re-identification of just one individual is considered a privacy breach and is subject to legal action and fines imposed by governments.

\subsubsection{Research}
De-anonymization attacks having such purpose intend to explore the weaknesses of published datasets and anonymization methods, advance the field of data privacy and encourage others to do so.
All publications cited in this work have such a purpose and most of them do not report a de-anonymization attack has been performed, but merely prove the de-anonimizability of the dataset(s) they study and report the extent to which it is de-anonymizable.
Those who report a de-anonymization attack has been performed, do not reveal the identities of the individuals they managed to re-identify and their findings are not confirmed by the dataset owners.

\subsubsection{Malicious}
An attacker with malicious intents would perform a de-anonymization attack in order to publicly reveal private, sensitive information of individuals who would, otherwise, not want to have this information published.

%\begin{itemize}
    %\item \textbf{Legal / GDPR aspect / Proof of Concept}:
    %Whether data could be de-anonymized.
    %De-anonymization is considered successful if just one individual is de-anonymized.
%    This aspect is not considered in the literature, but taken as granted.
 %   \item \textbf{Research}:
 %   All the references cited in this paper show that different kinds of data may de-anonymized and what portion of it, but they actually did not attempt de-anonymization or did not get their results of their de-anonymization attacks validated.
  %  \item \textbf{Malicious}:
   % Does not get scientifically published. (Might be articles however. I think I remember reading one.)
%\end{itemize}

\subsection{Target and Auxiliary Data}
The target dataset is defined as the anonymized dataset, potentially containing sensitive information of users that an attacker tries to de-anonymize.
The target dataset's availability might be public or private, with the public ones being the most used in research.
Some examples from the literature are:
health records \cite{Sweeney2000}, movie ratings \cite{Narayanan2008} and telecommunication mobility trace data \cite{Montjoye2013UniqueIT}.

The auxiliary data/information is defined as the non-anonymized information that aids the attacker in the de-anonymization of the target dataset.
The auxiliary information's availability might be public or private, with the public ones being the most used in research.
Some examples from the literature are:
voter registration list \cite{Sweeney2000}, IMDB's user reviews \cite{Narayanan2008} and social networks \cite{Srivatsa:2012:DMT:2382196.2382262}.
We categorize the auxiliary information into \textit{internal} and \textit{external}.

\textit{Internal} refers to the auxiliary information that comes from the same source as the target dataset.
This is the case of many research publications where the studied dataset is split into training and test set, corresponding to internal auxiliary information and target dataset, respectively.
Another case is when an attacker obtains part of the target dataset non-anonymized.

\textit{External} refers to the auxiliary information that does not come from the same source as the target data, and usually does not have the same structure as the target dataset.

%Define target data.
%\begin{itemize}
 %   \item \textbf{Public}:
  %  Most of the research cited here.
   % \item \textbf{Semi-public}:
%    That can be purchased.
 %   \item \textbf{Private}:
  %  A few in the literature, but still the de-anonymization attacks were not validated by the owner of the dataset.
%\end{itemize}
%Provide examples of kinds of datasets from the literature (or try to categorize them as well)....
%Define auxiliary dataset.
%\begin{itemize}
 %   \item \textbf{External Data}:
  %  Use an external not anonymized dataset(s) and try to match it with the target data.
   % \item \textbf{Training/Test Split}:
    %Split the target data into training and test sets and use the training as auxiliary and test as target.
%\end{itemize}
%Provide examples of kinds of datasets from the literature (or try to categorize them as well)....

\subsection{Approach}
The intuition behind all approaches in data de-anonymization is that all users have uniquely defining characteristics.
There exist varying approaches exploiting this uniqueness, whose complexity depends on the complexity of the data they apply to.
Therefore, in the following we will group the approaches by the type of data they apply to.

\textit{Low dimensional data} require the least complex approaches to de-anonymize.
Sweeney \cite{Sweeney2000} demonstrates that the combination of \{ZIP code, gender, date of birth\} is enough to uniquely identify 87\% of the U.S. population using the Generalized Dirichlet drawer principle (also known as the Pigeonhole principle).
Frankowski et. al \cite{Frankowski2006YouAW}, in one of their experiments, showed that even a simple set intersection is enough to identify a part of the users in a movie ratings dataset just by the movie mentions they made in another dataset.

\textit{High dimensional data} refers to transaction-like data which contain multiple interactions per user, such as ratings and web browsing history.
In order to de-anonymize such data, a model of the user must be built and the de-anonymization method usually depends on similarity between the users in the target dataset and the auxiliary information.
The intuition behind such methods is that each user's behavior is unique.
Narayanan and Shmatikov \cite{Narayanan2008} successfully matched up to 99\% of the users in the Netflix Prize dataset (movie ratings) to their publicly available IMDB profiles, by applying a scoring function.
This de-anonymization could reveal political, religious and societal views of the users that they do not publicly express, but are inferred through their private ratings of movies.
Frankowski et. al \cite{Frankowski2006YouAW} applied a similar approach for movie ratings and Su et. al \cite{Su2017DeanonymizingWB} for web browsing histories.

\textit{Graph data} are any kind of data that can be modelled with nodes connected with each other through edges.
The most notable cases are the social networks where users correspond to nodes and edges to the relationships between them (uni- or bi-directional).
The intuition here is that each user has a distinctive social network and, therefore, social graph.
Several approaches have been introduced exploiting the unique structure of the graphs, by trying to match the anonymized graphs to the non-anonymized ones \cite{Narayanan:2009:DSN:1607723.1608132}.

Some data can be categorized to either high dimensional or graph data depending on the way they are interpreted.
Trace data are such an example:
Gambs et. al \cite{Gambs2013} interpret mobility traces as a set of states and transitions and they define several metrics calculating the similarity between the anonymized and non-anonymized traces, while Srivatsa and Hicks \cite{Srivatsa:2012:DMT:2382196.2382262} interpret mobility traces as a graph consisting of meetings between users and match it to a non-anonymized social network graph.

%\textit{Trace data} are the kind of data that contain the location of users at specific points in time.
%The intuition behind the de-anonymization approaches on this kind of data is that users have a unique trail they follow.
%Similar methods to those of the high dimensional data have been proposed\cite{Gambs2013} and to those of the graph data as well\cite{Srivatsa:2012:DMT:2382196.2382262}.

%\subsection{Evaluation}
%How evaluation is performed.
%Categorize it as in Auxiliary Data.
%Describe how and provide metrics used in the literature.

\section{Discussion}
The intuitions of the de-anonymization approaches in the literature rely on the uniqueness of the users' characteristics and if an attack is to be performed on a large scale, a model needs to be built, which is, mostly, more complex than the Generalized Dirichlet drawer principle used by Sweeney to show that 87\% of the U.S. population is uniquely identifiable by the combination of \{ZIP code, gender, date of birth\} \cite{Sweeney2000}.
However, an average person with data processing skills would be interested in attacking very specific individuals and his/her attack would not require a complex model to be built and a naive approach might suffice.

Consider this example:
An attacker obtains access to an anonymized mobile phone trace dataset which contains the antenna a mobile phone is connected to at certain time intervals.
Would it be hard for the attacker to re-identify his/her neighbor in the dataset, if the attacker knows where the neighbor works or where he/she's been during the weekend?
Simple SQL queries to the dataset containing the location of an antenna and time of connection would most probably suffice.
As de Montjoye et. al \cite{Montjoye2013UniqueIT} claim:
\textit{"four spatio-temporal points are enough to uniquely identify 95\% of the individuals"} in such a dataset.

\section{Conclusion}
In this paper, we presented a taxonomy of the research in de-anonymization from an abstract point of view, oriented towards data providers.
Specifically, we identified three elements in the de-anonymization process, referenced prior research on de-anonymization and point out that de-anonymization is not as hard as one might think, if one is given strong auxiliary information.

%\section*{Acknowledgment}
%This research was supported by the Safe-DEED project (HORIZON 2020, Grant Agreement No.: 825225), funded by the European Commission. 

\nocite{*}
\bibliography{conference_idsc}{}
\bibliographystyle{IEEEtran}

\end{document}